\documentclass[sigconf]{acmart}

\copyrightyear{2020}
\acmYear{2020}
\setcopyright{licensedothergov}\acmConference[ICSEW'20]{IEEE/ACM 42nd International Conference on Software Engineering Workshops }{May 23--29, 2020}{Seoul, Republic of Korea}
\acmBooktitle{IEEE/ACM 42nd International Conference on Software Engineering Workshops (ICSEW'20), May 23--29, 2020, Seoul, Republic of Korea}
\acmPrice{15.00}
\acmDOI{10.1145/3387940.3392209}
\acmISBN{978-1-4503-7963-2/20/05}

\begin{document}

\title{A first look at an emerging model of community organizations
  for the long-term maintenance of ecosystems' packages}

\author{Th\'eo Zimmermann}
\email{theo@irif.fr}
\orcid{0002-3580-8806}
\affiliation{%
  \institution{Inria, Universit\'e de Paris, IRIF, CNRS}
  \postcode{F-75013}
  \city{Paris}
  \country{France}
}

\begin{abstract}
One of the biggest strength of many modern programming languages is
their rich open source package ecosystem.  Indeed, modern
language-specific package managers have made it much easier to share
reusable code and depend on components written by someone else (often
by total strangers).  However, while they make programmers more
productive, such practices create new health risks at the level of the
ecosystem: when a heavily-used package stops being maintained, all the
projects that depend on it are threatened.  In this paper, I ask three
questions.  \textbf{RQ1}: How prevalent is this threat?  In
particular, how many depended-upon packages are maintained by a single
person (who can drop out at any time)?  I show that this is the case
for a significant proportion of such packages.  \textbf{RQ2}: How can
project authors that depend on a package react to its maintainer
becoming unavailable?  I list a few options, and I focus in particular
on the notion of \emph{fork}.  \textbf{RQ3}: How can the programmers
of an ecosystem react collectively to such events, or prepare for
them?  I give a first look at an emerging model of community
organizations for the long-term maintenance of packages, that appeared
in several ecosystems.
\end{abstract}




\maketitle

\section{Introduction}

Efficient software engineering and the conception of large software
systems critically rely on reusable
code~\cite{Sametinger:1997:SER:260943}.  Since the rise of open source
ecosystems and package managers, software reuse has become a
systematic coding practice, even for small and medium-sized software
projects.  Decan \emph{et al.}~\cite{decan2019empirical} found a
majority of packages depend on other packages in all seven ecosystems
they studied.

Indeed, modern application-specific package managers have reduced the
cost of sharing pieces of reusable code (libraries) that a programmer
can extract from their project so that others can benefit from it.  It
has also made it much easier to rely on such a library in one's
project.  The consequence is that, nowadays, programmers will always
look for a package solving their problem before considering tackling
it by themselves.  The extreme case is the one of trivial
packages~\cite{abdalkareem2017developers}, which contain only one
(sometimes very straightforward) function, and yet still get many
users.

While creating and sharing a new package has now become very easy,
does not cost much to the package author, and may even be beneficial
to them,
%
maintaining a package long-term can be a real
burden~\cite{miller2019people}.  Furthermore, authors may have little
incentive to maintain some of their packages, considering that in some
cases, they are not even using them anymore~\cite{npm1}.
On the other hand, users that are currently using the library have
much more incentive to contribute to maintaining it, but this is not
always as easy for them as it is for the author.

The question of who should be responsible for maintaining an open
source library is therefore far from trivial.  And the answer may
actually depend a lot on the way the ecosystem is structured.

This paper begins by asking a first research question (\textbf{RQ1}):
how prevalent is the threat to ecosystems' health coming from packages
that are used by multiple projects but could suddenly stop being
maintained because a single person was responsible for them?

Next, we explore the various mitigation paths that projects can follow
when faced with such a situation (\textbf{RQ2}).

The paper concludes with a last research question (\textbf{RQ3}): what
can be done at the level of the ecosystem to mitigate such threats
when they occur, or reduce their chances of occuring in the future?
We will focus on a model of community organization, which emerged in
several ecosystems, that can simplify the process of hard forking and
provide a new home for unmaintained packages.

\section{Estimating the prevalence of single-maintainer packages}

\label{sec:single-maintainer-libraries}



The main issue of small libraries is generally that they have fewer
maintainers, most often a single maintainer, who might become less
active or outright missing for a number of reasons.  While the extreme
case of trivial packages~\cite{abdalkareem2017developers} can be
trivially solved by copy-pasting the code and maintaining it within a
larger source code base, I am interested in the frequent case of
non-trivial libraries that are worth keeping as separate dependencies
(to avoid duplicating innovation and maintenance work in the various
projects depending on them) but are still small enough to have a
single maintainer.

More specifically, I propose to focus on non-trivial libraries which
are depended upon directly by at least two actively maintained
software projects by distinct developers or developer teams: these
libraries are thus worth maintaining separately.  The goal of this
section is to estimate the proportion of such libraries which have a
single person maintaining them (i.e., a single developer has
push-access, even if changes might be contributed by others through
pull requests or e-mailed patches).  Given that it can happen that
maintainers of such libraries stop responding to requests and updating
their library for extended periods of time, I claim they represent a
very specific threat to the health of ecosystems.

This estimation will be based on the dataset provided by
Libraries.io~\cite{andrew_nesbitt_2017_808273} (cf. the first
companion Jupyter notebook~\cite{zimmermann2019librariesio}).  This
dataset includes a table of packages with source repositories,
containing more than 3,000,000 entries.
For this estimation, we focus on the 2,500,000 packages whose
repository is located on GitHub, because the dataset contains more
complete data for these.

A first filtering step of packages with more than one reverse
dependency and a repository size of at least 10 KB (in an attempt to
filter out trivial packages), makes this number go down to 257,000
packages, including 116,000 npm packages and between 10,000 and 30,000
PHP, Ruby, Python, and Maven packages.


To further filter only packages that are used in actively maintained
projects from different owners (individual developers or teams), we
extract the 800,000 GitHub repositories that were pushed to in the
last six months (before the Libraries.io dataset was published) from a
table which contains about 34,000,000 entries.

Finally, these data are joined by relying on Libraries.io's dependency
table, which includes about 390,000,000 entries.  This results in
about 65,000 packages that are depended upon by two actively
maintained projects from distinct owners.

Sometimes, many packages come from a single repository (more than
1,500 packages in the DefinitelyTyped repository%
\footnote{ DefinitelyTyped is a community repository gathering
  TypeScript type definitions for otherwise untyped Javascript
  packages: \url{http://definitelytyped.org/}.
	\label{footnote:definitelytyped}
}
and 286 packages in the Babel repository ---the next biggest
monolithic repository).  In this case, it is hard to estimate
different maintenance indexes for the different packages, and giving
the same index to thousands of packages would completely bias the
results, so we keep only one package per repository (the most depended
upon).

Out of the remaining 50,000 packages, \textbf{18\% have just a single
  contributor} (which is worse than having a single maintainer).

For each package owner, we can query GitHub to know
whether it is an organization and how many public members it has.
About \textbf{33\% of these 50,000 popular packages belong to
  organizations with at least two public members}.  Belonging to an
organization does not guarantee that the package will keep being
maintained, but it should prevent against a maintainer disappearing
without notice and no one having access to the repository.

For the rest of the package owners, I approximate the number of
maintainers by querying for the number of assignable users.
Assignable users are a super-set of collaborators with
write-access:%
\footnote{ For privacy reasons, GitHub does not share the list of
  collaborators or people with write-access on a repository, but it
  does share the list of assignable users nonetheless.  }
they correspond to organization members with read-access (when the
owner is an organization
%
) and collaborators that were manually added to the repository
itself~\cite{github_assign}.  \textbf{Of the remaining packages} that
were not part of an organization with two public members, \textbf{only
  33\% have two or more collaborators.}  This leaves a large
proportion of packages at risk.

\subsection{Threats to validity}

The goal of this section was to estimate the proportion of packages
that are sufficiently popular to be used in two maintained projects by
different owners, and yet have a single maintainer.  Two kinds of
errors could have affected the results: errors in the computation of
popular packages and errors in the computation of single-maintainer
packages.

Popular packages could have been missed, and in fact it is certain
that this was the case, since some popular packages are hosted outside
of GitHub.  Another possible error could have been with the
maintenance criterion.  Some projects should be considered maintained
even if they were not pushed to during a six-month period (some
feature-complete packages may need less frequent
updates~\cite{valiev2018ecosystem}, especially in a slowly evolving
ecosystem).  Furthermore, some projects may have been pushed to after
Libraries.io last refreshed GitHub's meta-data about them, and thus
the dataset may have contained outdated information that could have
resulted in marking these projects as unmaintained.  Therefore, the
number of maintained packages depending on a given package could have
been underestimated, and the package excluded because of this.
However, the resulting sample of popular packages is still rather
large (50,000 packages).  Bias in this sample could result mostly from
the exclusion of non-GitHub projects and of many packages hosted in
monolithic repositories.

I computed dependencies between packages irrespective of package
version numbers, so this could also lead to considering former
dependencies that were dropped or replaced.

Then, the criterion that was used to approximate the number of
maintainers is bound to have introduced some errors as well.  However,
I believe that it is more likely to have resulted in overestimating
the number of maintainers of a package, rather than underestimating
it.  Indeed, it is not because someone can be assigned issues in a
repository that they have access to all the required tools and
credentials to be considered an actual maintainer of the package.
Even if they are an actual maintainer, and are ready to take over from
the previous maintainer for a while, if they do not have admin-access,
they may not be able to nominate new maintainers, and this can harm
the maintenance of the package after some time, in particular if they
want to step down eventually.

Overall, I believe that these threats do not put much doubt on the
main conclusion of my analysis which is that many popular packages
have a single maintainer.

\subsection{Related work}

\subsubsection{Project maintainers and developers}

Yamashita \emph{et al.}~\cite{yamashita2015revisiting} analyzed the
proportion of core developers in 2,496 GitHub projects.  They used
various criteria to define core developers, one of them using GitHub's
collaborator API endpoint (through the GHTorrent
dataset~\cite{gousios2012ghtorrent}).  Unfortunately, access to this
API has been restricted since then and GHTorrent does not include this
data anymore.
%
This is why I used the assignable users information instead.

My analysis is different from studies computing metrics such as
\emph{bus factor} (also called \emph{truck
  factor})~\cite{ferreira2019algorithms,torchiano2011my}, and more
generally estimating turnover
risks~\cite{rigby2016quantifying,nassif2017revisiting} within a
software project, because I am focusing on the problem of who is able
to integrate changes and publish new versions of a package.  It can
happen that popular packages receiving pull requests from many
contributors still have a single maintainer that suddenly disappears.
On the other hand, a project that is not subject to such risks,
because it belongs to an active organization or has several
maintainers, can still be subject to knowledge loss risks.  Therefore,
the two types of studies are complementary, and highlight different,
but related, health risks that can affect a package ecosystem.

Avelino \emph{et al.}~\cite{avelino2019abandonment} have found a
significant proportion of projects having faced the event of a
``truck-factor'' developer stepping down, despite having only analyzed
about 1,900 popular repositories.  Less than half of the projects
survived, generally because a new or preexisting contributor took
over.  They interviewed these contributors that helped projects
survive, and identified the difficulty of getting access to the
repository as a significant barrier (when project maintainers had
become unresponsive).

\subsubsection{Unmaintained projects}

Several studies have highlighted the fact that many open source
projects are dormant or
abandoned~\cite{khondhu2013all,kalliamvakou2016depth}.  However,
a project threatening to become dormant does not pose the same risk to
open source ecosystems depending on whether it has a lot of users,
very few, or none beyond its author.

Valiev \emph{et al.}~\cite{valiev2018ecosystem} have found clear
evidence that packages that have been able to gather a large community
of users over time
are much less likely to become dormant.  This expected result does not
contradict the observation that this risk is still present and strong
for a number of popular packages.

\subsubsection{Ecosystem health}

Measuring ecosystem health is an important and active research
question~\cite{jansen2014measuring,mens2017towards}.  My work relates
to this literature by highlighting a health risk factor
(single-maintainer projects) that could be integrated in health
assessment frameworks.

\section{Coping with unmaintained packages}

When a package becomes unmaintained (the maintainer does not respond
anymore to issues and pull requests and does not push new commits or
versions), what can the projects using it do?

\subsection{Removing or replacing the dependency}

Upon discovering that a project depends on an unhealthy dependency, it
can be time to reevaluate the usefulness of the latter.  Sometimes,
the functionality brought by the dependency is not that useful, or an
alternative, healthier package could be used instead.  Removing or
replacing the dependency can, in such case, turn out to be beneficial,
although the migration can bear significant costs to the project, not
necessarily at the best of time.

Furthermore, this is a solution that each project has to evaluate on
their own, and while it might be feasible for some projects to drop
the dependency, it might be significantly harder for others.  Having
many projects migrate away from an unhealthy library can further
reduce its chances of surviving, thus threaten other projects for
which migrating is too costly.

Previous studies have mined data from projects having performed
library migrations to suggest candidate libraries to migrate
to~\cite{teyton2012mining} and to map between methods of two
libraries~\cite{teyton2013automatic,alrubaye2019migration}.

\subsection{Vendoring}

Vendoring a dependency is the process of copying its sources within
the project's sources and building the whole thing, instead of
installing the dependency with a package manager first and building
the project by relying on the installed dependency.


Vendoring dependencies allows integrating patches proactively.
Some of these patches might have been found in unmerged pull requests
from external contributors.  However, this solution cannot be a
long-term solution because it is more work for everyone to have to
integrate patches manually, and at some point new pull requests cannot
continue to be based on an unchanged base branch.

\subsection{Forking}

\subsubsection{Definition}

Forking has a broad meaning today in open source.

Early academic works which studied forking considered only what is
usually denoted today as \emph{hard forks}.  For instance, in their
2012 paper~\cite{robles2012comprehensive}, Robles and
Gonz\'alez-Barahona give a definition of a fork that includes
requirements such as having a new project name and a disjoint
community.  The Hacker's Dictionary~\cite{raymond1996new} even
specifies that the two code bases must be developed in parallel and
have irreconcilable differences between them.

The ``right to fork'' is qualified by
Weber~\cite[page~64]{weber2004success} as an essential freedom of free
software.
Nyman and Lindman~\cite{nyman2013code} claim that forking is the most
important tool to guarantee sustainability in open source development,
and that the right to fork has a major effect on governance, even in
the absence of any forks.

With the rise of GitHub, forking has taken a new meaning.
\emph{Development forks}~\cite[Chapter~8]{fogel2005producing} are
copies of the sources where a contributor makes changes to the code in
order to submit them for review through the pull request mechanism.

But even before GitHub, forking was much more common and much less
definitive than hackers and researchers alike seemed to believe.
Nyman and Mikkonen~\cite{nyman2011fork} observed the presence of many
forks on SourceForge, including forks claimed to be temporary and
hoping to get their changes integrated upstream (that can therefore be
classified as development forks).  They also noticed the phenomenon of
forking a project because it seemed to be abandoned, and not because
of some disagreement.  While this should still be denoted as a hard
fork, there is only one project under development after the fork,
contrary to the definition of the Hacker's
Dictionary~\cite{raymond1996new}.  This is the kind of forks that we
are interested in, in this section.  We denote this sub-type of hard
forks as \emph{friendly forks}.

\subsubsection{Socio-technical issues when forking a package}

\paragraph{When to advertise a friendly fork?}

While it is easy for anyone to maintain a personal fork of a project,
which contains the original code with some modifications on top, it
may be difficult to decide when to advertise this project as a
friendly fork intending to take over the place of the original,
unmaintained project.

First, maintaining a fork of an unmaintained project for a long time
without doing any advertisement is likely to result in duplicated
work, as other persons interested in the project, but who are not
aware of the fork, run in the same issues and prepare their own fixes.
Zhou \emph{et al.}~\cite{zhou2019fork} studied inefficiencies that may
arise mainly due to lack of awareness of the work that was done in
forks.

Besides, putting efforts into advertising a fork
can pay back by bringing an influx of new contributions from
developers that were interested in the project but were discouraged by
the absence of feedback from the maintainer.  Still, it requires time
and commitment.
The model of community forks that is discussed in
Section~\ref{sec:community} can help reduce the level of commitment
required.

The time to wait until the source project is considered unmaintained
can also vary depending on community expectations, and is rarely clear
to anyone.  While SourceForge's ``Abandoned Project Takeover'' page
set a 90-day delay to get an answer,%
\footnote{Cf. \url{https://web.archive.org/web/20100609115535/http://sourceforge.net/apps/trac/sourceforge/wiki/Abandoned\%20Project\%20Takeovers}}
CPAN's FAQ~\cite{cpan_faq} informally sets a delay of one year without
response before considering transferring maintenance of a module.

\paragraph{How to fork on GitHub?}

On GitHub, forking a project is as easy as clicking on a button.  But,
when preparing a hard fork,
the new maintainer may wonder whether this is the right choice.

By default, forks on GitHub are not meant to take over a project:
issues are disabled (but they can easily be switched on) and a
prominent link to the source project is displayed under the project's
name.  Besides, code is not searchable in a fork unless it has more
stars than its source~\cite{github_search_forks} (which can take quite
some time to get).

GitHub does not make discovering maintained forks very easy: the only
way to learn about them is to display the fork tree, which is often
very large.
When the forks are too numerous, GitHub will not display the full list
and the most important ones may be missing.
The ``Lovely forks'' browser extension~\cite{upadhyay_forks} helps
developers discover notable forks by querying for them and showing
them prominently, where GitHub would display a fork's source.

An alternative solution is to create a new repository manually, and to
push the content of the original repository in it.  It is also
possible to contact GitHub staff to remove the fork status from an
existing repository.  They can also change the base directory in a
fork network, but this requires consent from the original
owner.%
\footnote{ As I was told in a private mail by a member of GitHub
  staff. }

\paragraph{Migrating issues and pull requests.}

GitHub does not provide any support for easily duplicating issues
between two repositories.  Doing so is nonetheless possible using a
tool such as github2github~\cite{github2github}.  Reusing the exact
same numbers for imported issues is technically feasible.
The advantage of doing this is that the code and the commit message
frequently reference issues by their number, and importing preexisting
issues ensures that these numbers continue to make sense.  On the
other hand, new maintainers might appreciate the ability to duplicate
only a subset of issues they intend to solve.

If the fork was created using GitHub's fork button, it is also
possible to manually recreate pull requests for every pull request
that is still opened on the original repository.

\paragraph{How to publish updates to the package registry?}

Different registries and different ecosystems have different views
regarding the transfer of a package to a new maintainer.  Most support
voluntary transfers, and some also support transfer to a new
maintainer when the previous maintainer is completely unresponsive.

In some registries, all packages are scoped (the package name is
prefixed by the name of the author), so unless an author explicitly
gives access to a new maintainer, there is no way to continue using
the package full name, and everyone will have to update their
dependencies.  On the other hand, it means the source project does not
get a special status in the package index compared to its forks.

In some registries that support both non-scoped and scoped package
names, keeping the same base name while adding a scope can be a way of
marking the affiliation of the package to the original one.%
\footnote{ As recommended in this Open Source Stack Exchange answer
  about npm: \url{https://opensource.stackexchange.com/a/7025/5858} }
This is also the technique that is used by the DefinitelyTyped
repository (cf. Footnote~\ref{footnote:definitelytyped}) to publish
type definitions corresponding to untyped Javascript packages.

In registries based on a shared repository of manifest files, it is
technically easy to change the source of a package when publishing an
update.  The question of whether it gets accepted will depend on the
registry's policy. For instance, MELPA's
policy~\cite{melpa_contributing} says that forks will not be accepted
except in ``extreme circumstances''.

Early archives where sources (and sometimes even bug trackers) are
located on the platform make it technically even easier to change the
maintainer of a package: CTAN, CPAN, CRAN and PEAR all have
documentation regarding unmaintained / orphaned packages.  CTAN
specifies that modifications to a package should come from the package
author or maintainer, new maintainers can be accredited by the current
maintainer, but leaves the door open to discussing a solution when a
package is unmaintained and the author is
unresponsive~\cite{ctan_orphaned}.  CPAN administrators can transfer
maintenance of a package to a new volunteer after sufficiently many
steps have been taken to reach the previous maintainer and advertise
the intent to take over the package~\cite{cpan_faq}.  CRAN has a
formal orphaning process, after which volunteers can request to become
the new maintainers~\cite{cran_policy}.  For non-orphaned packages,
transfer requires written agreement of the previous maintainer.  PEAR
packages can be marked as unmaintained, and may then be transferred to
a new lead maintainer~\cite{pear_orphaned}.

In general, there is a trust issue associated to allowing a change of
maintainer for a package (as this means that someone can update a
dependency to a new version without realizing the change of
ownership), but this pertains to a much more general trust question in
code reuse and package ecosystems.  Because of this issue, npm is even
considering restricting the rules for voluntary
transfers~\cite{npm-transfer}.

We can see that forking, while often the best solution for the user
community, puts a very large cost on the new maintainer when they do
things right and try to organize the community around the new fork.
And when forks are created but not properly advertised, it can only
lead to duplication of effort.  In the next section, I present a
solution that alleviates the cost of forking.

\section{Ecosystem-level solutions}

\label{sec:community}

\subsection{Community forks to increase sustainability}

While hard forks are a possible solution to the problem of
unmaintained packages, we have seen that this solution puts a
significant cost on the new maintainer.  It can also put a significant
cost on the user community when the package registry does not make it
easy to switch the maintainer of a package, because a possibly large
number of users will need to learn about the hard fork, evaluate if it
is likely to be viable, and update their dependencies.

The transition period, which starts when people become aware that the
previous maintainer has disappeared and only ends when the user
community has massively adopted a hard fork,
is a time during which it is likely that efforts are duplicated,
potential contributions as pull requests or bug reports are wasted
because they are being ignored or users stop submitting them, the
package user base stops growing, or even shrinks, as people look for
alternatives to migrate to, or even start their own from scratch, etc.

This cost can be deemed too high, especially if the hard fork itself
also has a single maintainer, and thus risks suffering from the same
issues a few years later.  It is natural that the user community can
anticipate this and will be reluctant to move massively to a hard fork
that does not take steps to prevent this scenario to repeat.

A preventive measure would typically be the creation of a
\emph{community fork}.  When a package raises sufficient interest and
enough people are motivated to keep maintaining it together, they can
host the new repository in a dedicated GitHub organization instead of
a personal account and ensure that, at any time, there will be several
administrators of this organization and several persons with
credentials to publish a new version of the package.

However, most popular single-maintainer packages are likely to be too
small for such a community fork to happen.  To facilitate the creation
of community forks, a possible model is to host them in a community
organization dedicated to the long-term maintenance of important
packages.  This is the model that I present now.

\subsection{A model of community organizations}

\label{sec:community-org-model}

In this model, a single informal organization is created (typically as
an organizational account on GitHub)
to host community forks in a specific ecosystem (typically around a
programming language or framework).  A place is dedicated to
discussing organizational aspects of the community and to proposing
new packages for inclusion (for instance the issue tracker of a
meta-repository).  The criteria for accepting a package may vary, but
generally include having at least one person who volunteers to
maintain the package.

Maintenance may actually be a community effort, but the advantage of
having a designated maintainer for a package is to avoid diluting
responsibility.  However, volunteering to be the principal maintainer
of a package is not a long-term commitment.  The point of hosting
packages within a community organization is that the maintenance
responsibility can easily be transferred if a maintainer wants to step
down or becomes unresponsive, as long as there are responsive
organization owners and a new volunteer maintainer.

Such community organizations can also facilitate collaboration and
encourage maintainers to share best practices.  If all maintainers are
given commit access to all projects, one can easily help with another
package while its own maintainer is temporarily unavailable.

Hosting hard forks in such community organizations is likely to be a
factor that will help users adopt them faster, as it guarantees
against the risks associated with a new single-maintainer package.

Finally, the existence of such community organizations provides an
exit strategy for authors of popular packages that would like to step
down from maintaining them.  They can submit their package for
inclusion and transfer them to the community if accepted.  Again,
inclusion criteria may vary depending on the specific organization.

\subsection{The case of elm-community}

One early instance of this model
%
is elm-community ({\small \url{https://github.com/elm-community}}),
which was founded in November 2015.  On July 5\textsuperscript{th},
2019, I interviewed Ryan Rempel, the founder of the organization, who
explained to me how this organization came about.

The Elm package ecosystem had already got a culture of package forking
and updating every time a new version of the Elm language was
published and some original package authors failed to react.  This was
made easier by the fact that all package names are scoped in the Elm
package registry.  Consequently, the source package does not have a
special status compared to its forks.

However, some impure Elm packages (containing what people used to call
native code and now call kernel code~\cite{czaplicki2018native},
i.e., JavaScript) had to go through a formal ``blessing''
(whitelisting) process to be published in the Elm package registry.
For such packages, forking and updating could not be done so casually,
because it would additionally require going through this formal
approval process.  This specific issue was discussed on the Elm users'
mailing list%
\footnote{ Cf.
  \url{https://groups.google.com/forum/\#!topic/elm-discuss/-GQJkWGdMvg/discussion}
}
after the author of a widely used package, containing native code, had
been unresponsive for two months.  During that time, a pull request
with a trivial patch required to upgrade the package to the new
version of the language (Elm 0.16) had been left unanswered.

Max Goldstein, an active community member who is now part of Elm's
core team, suggested the creation of an ``elm-community'' GitHub
organization ``to steward the most important non-official packages''.
Ryan Rempel, another active community member and the author of the
unmerged pull request, jumped on the idea, created the organization,
forked the package, and submitted a whitelisting request for it, on
the same day.  Two days later, he created a meta-repository named
``Manifesto'' in which he described the purpose of the organization in
a README and whose issue tracker served to host organizational
discussions and package adoption requests.

Ryan told me that the reason he had reacted so quickly after the idea
was first proposed was because he had viewed this as an opportunity to
foster a new form of collaboration, that would be less disciplined and
less centrally controlled than what was common in the Elm community.
Indeed, he told me, the Elm community is unusually disciplined for an
open source community, around a core team that has very specific ideas
about what kind of participation is welcome.  All his previous
attempts to advocate a more open community had failed, and left him
tired.  So in this case, he started the community organization without
really leaving any time to anyone to discuss the idea, invited five or
six people from the beginning, and it turned out to be successful
pretty quickly.

Shortly after, it gained some repositories that made more sense to
develop collaboratively, e.g., a series of standard library
extensions.%
\footnote{ Cf.
  \url{https://groups.google.com/forum/\#!topic/elm-discuss/wJPvZUql6v0/discussion}
}

The creation of a Manifesto repository was initially meant as a way of
explaining the philosophy of the project, but also to allow issues to
be used for organizational questions.  The word ``Manifesto'' was
slightly political, but the text of the README avoided any provocative
content.  According to Ryan, the text was rather abstract at the
beginning, and others helped make it more concrete over time.  The
governance, in particular, remained voluntarily informal.

The idea to have a principal maintainer for each repository to avoid
dilution of responsibility (which leads to issues and pull requests
being left unanswered) was introduced about six months later by a
member of the organization.%
\footnote{Cf.
  \url{https://github.com/elm-community/Manifesto/issues/16} }
This change made explicit the rule that issues and pull requests are
normally handled by the repository's principal maintainer, but in case
of unresponsiveness, any other member can step in, and in case of
long-term unresponsiveness, the maintainer is changed.

\subsection{An emerging model}

\label{sec:community-emerging}

Elm-community is not the only, nor the first, instance of this model.

Vox Pupuli ({\small \url{https://voxpupuli.org}}) was founded in
September 2014 to maintain Puppet modules (initially under the name of
puppet-community) and currently hosts over 176 repositories and 139
collaborators.  They have precise migration
documentation~\cite{voxpupuli2016migrating} which clearly states a
preference for repository transfer, but supports hard forks when
package authors are completely unresponsive.  They have also made
efforts in recent years to promote their model so that other
communities can inspire from it to create their
own~\cite{galic2017voxpupuli,hollinger2017voxpupuli}.

Sous Chefs ({\small \url{https://sous-chefs.org/}}) was founded in May
2015%
\footnote{Cf.
  \url{http://lists.opscode.com/sympa/arc/chef/2015-05/msg00091.html}
}
to maintain Chefs ``cookbooks'' (initially under the name of the Chef
Brigade) and had a meta-repository from the start.%
\footnote{ The first issue
  (\url{https://github.com/sous-chefs/meta/issues/1}) and the first
  commit (\texttt{efb426f}) were created three days after the initial
  mailing list announcement.  }
They also have a clearly documented forking and transfer
policy~\cite{souschefs_fork,souschefs_transfer}.  In particular, their
forking policy states that hard forks are republished to the package
registry under the original name with a ``sc-'' prefix.

Both organizations are pretty open to any repository transfer from
members of the organization.

DLang-community ({\small \url{https://github.com/dlang-community}}),
founded in December 2016, has stricter inclusion criteria than Vox
Pupuli or Sous Chefs: packages are not transferred or created in the
organization without it being discussed with other members, and the
package being important to the community.

The following are direct or indirect elm-community descendants:
\begin{itemize}
	\item ReasonML-community ({\small
          \url{https://github.com/reasonml-community}}) was founded in
          January 2017 under the name Buckletypes (inspired by the
          DefinitelyTyped repository,
          cf. Footnote~\ref{footnote:definitelytyped}).  It was
          renamed in July 2017, and got a meta-repository influenced
          by elm-community in January 2018.\footnote{ See
            \url{https://github.com/glennsl/reasonml-community-meta-proposal}
            and
            \url{https://github.com/reasonml-community/meta/issues/1}.
          } However, it has failed to get clear adoption guidelines,
          its meta-repository is almost unused, and it has recently
          taken almost a full year to host a fork of the very popular
          graphql\_ppx package after its author had stopped
          responding.
	\item Coq-community ({\small
          \url{https://github.com/coq-community}}), founded in July
          2018, was directly influenced by elm-community.
	\item OCaml-community ({\small
          \url{https://github.com/ocaml-community}}), \\ founded in
          August 2018, was influenced by coq-community and
          elm-community.  Similarly to DLang-community, it only
          accepts popular OCaml packages.
	\item React-native-community
          (\url{https://github.com/react-native-community}) was
          founded in July 2016, but got a ``renaissance'' period
          starting in December 2018 that was influenced by
          OCaml-community.\footnote{Cf.  {\fontsize{6.5}{6.5}
              \url{https://github.com/react-native-community/discussions-and-proposals/issues/63}}
          }
\end{itemize}

Some organizations are similarly structured and intend to ease package
maintenance after their maintainer has left or lost interest but do
not explicitly support hard forks.  Examples include the F\# Community
Incubation space ({\small
  \url{https://github.com/fsprojects/FsProjectsAdmin}}), Electron
Userland (\url{https://github.com/electron-userland}), the Elytra
group ({\small \url{https://github.com/elytra}}) which does not have a
meta-repository or general guidelines, but does list the maintainers
of each repository in their description, and advertise when a module
is looking for a new maintainer.  The Fluent Plugins Nursery ({\small
  \url{https://github.com/fluent-plugins-nursery}}) is explicitly
intended for plugins that are not maintained by their original author
but also states ``we don't want to fork original authors'~''.

\paragraph{Identifying community organizations}

In the following, I present a process I used to discover such
organizations (cf. the second companion Jupyter
notebook~\cite{zimmermann2019community}).

First, I listed 75 keywords that could be expected to appear in the
name or the description of such organizations.  This included keywords
expressing collaborative work such as
``collective'',
``maintain'', ``participate'',
but also keywords expressing what is being worked on such as
``library'', ``module'', ``package''.

I used GitHub's search, via the GraphQL API~\cite{github_graphql_API},
to query for organizations which matched one of these keywords, and
which had at least 5 repositories.  GitHub's search only gives access
to 1000 results, so when the number of results was above this limit, I
further split the search using language filters.
This first step yielded over 30,000 organizations (this is close to
15\% of all GitHub organizations with at least 5 repositories).

The second step consisted in applying some filters on the results.
Since I was only interested in community organizations, I filtered out
the ones that had less than 10 public members and less than 10
assignable users on the most starred repository (as a way of
estimating the number of collaborators for organizations with mostly
private members).  Since I was interested in organizations maintaining
important packages, I also filtered out the ones whose most popular
repository had less than 10 stars.

The third step was intended to further reduce the list to
organizations that have received repository transfers.  Unfortunately,
GitHub does not give access to this information.  I used a trick which
consists in comparing the creation date of the organization to the
creation date of its repositories.  If an organization contains
repositories that predate its creation, they have necessarily been
transferred.  Obviously, the converse is not true, so it could have
resulted in underestimating the number of transferred repositories.


I manually browsed the resulting table of 938 organizations with at
least two such transferred repositories.  I used the name and
description of the organization to infer its purpose.  I eliminated
many organizations whose description was something like ``community
packages for X'' when the organization's website was actually the
website of X.  Indeed, this case is too frequent, and the reuse of the
main product's website is a good indicator of the absence of a website
or meta-repository specific to the community organization.

When an organization seemed to correspond to the type I was searching,
I opened its website or GitHub page and looked for more information
about it.  It was frequent to find organizations with many
repositories but no information on its principles and whether it
accepted new members or new projects.

I am well aware that applying this series of filters is very likely to
have resulted in missing out organizations that still fit the model I
presented in Section~\ref{sec:community-org-model}.  Nevertheless, the
number of examples that I found, and the absence of relationship
between many of them (in particular, between elm-community and the two
organizations that predate it) leads me to think that \textbf{this
  model of organization is naturally emerging in open source package
  ecosystems} in answer to the recurring problem of single-maintainer
libraries that I presented in
Section~\ref{sec:single-maintainer-libraries}.  The fact that such
organizations frequently got inspiration from one another shows that,
while the need for such an organization is natural, the exact way of
structuring it is not.  Therefore, this contribution is important
because, by surveying existing instances, we can bring useful
information to practitioners wondering about the opportunity of
founding such organizations, and how to structure them.

\subsection{How does it compare to earlier models?}

The idea of shared infrastructure to develop several projects together
is not new.  For instance, PEAR~\cite{pear} was PHP's first package
registry, but it was also much more.  Packages had to go through a
formal submission process to get accepted.  Once they were, they got a
repository and bug tracker.  Finally, when a package was left
unmaintained by its author, a new maintainer could be
appointed~\cite{pear_orphaned}.

However, there was an important difference in the motivations of
authors applying to get their packages accepted into PEAR compared to
authors, or interested users, proposing a package to a community
organization today.  As PEAR was PHP's only package manager and
registry at the time (until support for alternative registries was
introduced, starting in 2005\footnote{
  Cf. \url{http://php-pear.1086190.n5.nabble.com/Questions-about-PEAR-amp-Composer-td36854.html}.
}), authors submitted their packages to get it distributed, rather
than for the shared maintenance model.  For some maintainers, the
model and infrastructure were convenient, but others preferred to host
their projects elsewhere.



Another well-known example is the Apache Foundation, which hosts many
open source projects and provides shared processes and infrastructure.
The main difference with the model of community organization presented
in this paper is that the Apache Foundation only accepts larger
projects (all the projects have several maintainers and their own
mailing lists~\cite{apache_how_it_works}).  The Apache Foundation
model is therefore not suited to solve the issue of single-maintainer
packages, and does not support hard forking (despite the Apache web
server being itself a community fork of an unmaintained project).

\subsection{Open issues, future work}

\label{sec:open-issues-community-org}

I have presented a model of community organization for the
collaborative, long-term maintenance of an ecosystem's packages, and I
have identified numerous instances of this model.  However, the method
I used to find these examples is not exhaustive.

The use of GitHub's search to list possible candidates is good for
exploratory work, but cannot be used beyond that: I have observed that
the results obtained are very unstable when trying to fetch them
again, and I have also found a number of bugs in GitHub's search
filters.  According to GitHub staff, this is because the search index
is sometimes out of date.

Finding more examples will require coming up with more precise
criteria to detect automatically this kind of organizations, probably
starting from the complete list of about 2,000,000 GitHub
organizations.  Then, we could try to identify which characteristics
of an ecosystem favor the emergence of such organizations.

Given that this is an emerging model, each instance is unique and it
would be useful to come up with a number of parameters that can be
used to describe them.  Interviewing founders and participants would
be helpful in that regard.  Then, the next step would be to identify
which of these parameters are associated with successful community
organizations.  For instance, it would seem that documenting the
adoption process is helpful to ensure that people know what to do when
a useful package in the ecosystem has been abandoned and would be a
candidate for adoption, whereas the absence of such guidelines can
lead to a lack of reactivity and duplication of effort with multiple
people starting forks (as happened with the graphql\_ppx library in
the ReasonML ecosystem).

Finally, it seems important to systematically assess the impact of
such community organizations on an ecosystem and, in particular, on
the packages that get transferred or forked in the organization.  For
instance, Zhou \emph{et al.}~\cite{zhou2019fork} studied
inefficiencies in fork-based development, such as duplication of work
or community fragmentation.  We could try to evaluate whether the
presence of a community organization helps reduce these
inefficiencies, and under which conditions.  This would provide
concrete incentives to practitioners to create more instances of this
model.

\section{Conclusion}

Modern and attractive package ecosystems are made of a multitude of
small and large open source libraries.  When a popular package is
depended upon by many projects but has only a single maintainer, it
creates a risk of the maintainer suddenly becoming unresponsive and
the package not being updated anymore.  I have shown that a large
proportion of popular packages are single-maintainer packages and I
have presented some mitigation paths that users of these packages can
follow in such situations of unresponsiveness.  Among them, the
creation of a friendly fork is often the best for the community, but
it can be costly to start a fork, and it does not help very much if
the new fork is also a single-maintainer package.  This is why users
can create community organizations for the collaborative, long-term
maintenance of an ecosystem's packages.  Such organizations can reduce
the cost of forking, while creating better guarantees for the future
of the fork.  I have shown that this model of community organizations
has emerged in a number of package ecosystems and, as a first step
toward a more systematic analysis, I have presented the case of
elm-community in greater details.

\bibliographystyle{ACM-Reference-Format}
\bibliography{biblio}

\end{document}